\documentclass[aps, prb, showpacs,amsmath,amssymb,amsfonts,10pt
	,onecolumn
	,superscriptaddress,floatfix,footinbib
	,reprint
	]{revtex4-1}

\usepackage[utf8]{inputenc}
\usepackage{amsmath}
\usepackage{amssymb}
\usepackage{graphicx}

\usepackage{upgreek}
\usepackage{SIunits}
\usepackage{dsfont}

\usepackage{units}
\usepackage{color}

\usepackage{tabularx}

\begin{document}


\title{Dipole Analysis of the Dielectric Function of Colour Dispersive Materials: \\Application to Monoclinic Ga$_2$O$_3$}

\author{C. Sturm}
\affiliation{Institut f\"ur Experimentelle Physik II, Universit\"at Leipzig, Linnéstr. 5, 04103 Leipzig, Germany}

\author{R. Schmidt-Grund}
\affiliation{Institut f\"ur Experimentelle Physik II, Universit\"at Leipzig, Linnéstr. 5, 04103 Leipzig, Germany}

\author{C. Kranert}
\affiliation{Institut f\"ur Experimentelle Physik II, Universit\"at Leipzig, Linnéstr. 5, 04103 Leipzig, Germany}

 \author{J. Furthm\"uller}
 \affiliation{Institut f\"ur Festk\"orpertheorie und -optik, Friedrich-Schiller-Universit\"at Jena, Max-Wien-Platz 1, 07743 Jena, Germany}


\author{F. Bechstedt}
 \affiliation{Institut f\"ur Festk\"orpertheorie und -optik, Friedrich-Schiller-Universit\"at Jena, Max-Wien-Platz 1, 07743 Jena, Germany}

\author{M. Grundmann}
\affiliation{Institut f\"ur Experimentelle Physik II, Universit\"at Leipzig, Linnéstr. 5, 04103 Leipzig, Germany}

\begin{abstract}
We apply a generalized model for the determination and analysis of the dielectric function 
of optically anisotropic materials with colour dispersion to phonon 
modes and show that it can also be generalized to excitonic polarizabilities and 
electronic band-band transitions. We take into account that the tensor 
components of the dielectric function within the cartesian coordinate system 
are not independent from each 
other but are rather projections of the polarization of dipoles oscillating 
along directions defined by the, non-cartesian, crystal symmetry and 
polarizability. The dielectric function is then composed of a series of 
oscillators pointing in different directions. The application of this model is exemplarily demonstrated for monoclinic ($\beta$-phase) Ga$_2$O$_3$ bulk single crystals. Using this model, we are able to relate electronic transitions observed in the dielectric function to atomic bond directions and orbitals in the real space crystal structure. For thin films revealing rotational domains we show that the optical biaxiality is reduced to uniaxial optical response.
\end{abstract}

\maketitle

\section{Introduction}

For the understanding, design and fabrication of optoelectronic devices, the optical properties of the involved materials have to be known. A well established and powerful 
method for the determination of these properties is spectroscopic 
ellipsometry \cite{AzzamBashara, Fujiwara}. We concentrate here on the 
dielectric function (DF), which is usually obtained by means of numerical model 
analysis of the experimental ellipsometry data and then often described by a 
series of line-shape model dielectric functions in order to deduce phonon 
properties, free charge carrier concentrations and the properties of electronic 
transitions (e.g Ref.~\onlinecite{Schubert2010, Fujiwara}). For isotropic 
materials this method is well established. However, in recent years, 
optically anisotropic materials, as e.g. Ga$_2$O$_3$ \cite{Ueda1997, 
Yamaguchi2004, Varley2015, Sturm2015}, CdWO$_4$ \cite{Jellison2011} and 
lutetium oxyorthosilicate \cite{Jellison2012}, went into focus of research 
since they are promising candidates for optoelectronic applications in the UV 
spectral range. However, the determination of their optical and electronic 
properties is more challenging compared to isotropic materials since they 
depend on the crystal orientation. The dielectric function is represented by a (frequency-dependent) tensor and the determination 
of its components requires a series of measurements for various crystal 
orientations.

For (non-chiral) optically anisotropic materials, the dielectric function is in general a symmetric tensor consisting of six independent components \cite{BornWolf}, i.e.
\begin{equation}
	\varepsilon = 
	\begin{pmatrix}
		\varepsilon_{xx} & \varepsilon_{xy} & \varepsilon_{xz} \\ 
		\varepsilon_{xy} & \varepsilon_{yy} & \varepsilon_{yz} \\
		\varepsilon_{xz} & \varepsilon_{yz} & \varepsilon_{zz} 
	\end{pmatrix}\,.
	\label{eq:df_tensor_general}
\end{equation}
Due to its symmetry, this tensor can be diagonalized independently for the real 
and imaginary part at each wavelength separately. In the transparent spectral 
range, i.e. for vanishing imaginary part, the diagonal elements are the 
semi-principal axes of the ellipsoid of wave normals and are often called 
dielectric axes. For materials with monoclinic or triclinic 
crystal structure, the orientation of the dielectric axes depends on the 
wavelength and is often called colour dispersion. In the spectral 
rang with non-vanishing absorption, the situation becomes even more 
complex. Due to the independent diagonalizability of the tensor
(\ref{eq:df_tensor_general}) for the real 
and imaginary part, the corresponding dielectric axes in general do not 
coincide which 
each other. Thus, in general, four dielectric axes are present. For these 
classes of materials only few reports on the determination of the 
full dielectric tensor exists, e.g. for $\alpha$-PTCDA \cite{Alonso2002}, 
pentacene \cite{Dressel2008}, BiFeO$_3$ \cite{Schmidt2015}, CdWO$_4$ 
\cite{Jellison2011}, K$_2$Cr$_2$O$_7$ \cite{Hoefer2014}, CuSO$_4 \cdot 5$H$_2$O 
\cite{Hoefer2013} and effective anisotropic materials as e.g. slanted columnar 
films 
\cite{Schmidt2013}. Most of these works are limited to the 
determination of the line shape of the dielectric function, treating each 
tensor component of the DF independently of each other. This can, from a 
technical point of view, result in large correlations between the individual 
tensor elements causing non-physical results. More importantly, the deeper nature of 
polarizabilities in the material, like phonons, excitons, and 
electronic band-band transitions, cannot be explored this way. Thus,
lineshape model dielectic functions (MDF) representing the oscillators properties like energy, amplitude, broadening, and even oscillation direction in a meaningful and physical correct way have to be used.

Facing this, Dressel \textit{et al.} \cite{Dressel2008} proposed an approach 
assuming that the dipole moments are aligned to three polarization axis which 
should coincide with the crystallographic axes. Taking this model into account, 
the dielectric tensor is fully described by its three independent principal 
elements and the known angles between the crystallographic axis. However, as a 
consequence of this approach the principal axes of the indicatrix (related 
to the real part of ${\varepsilon}$) coincide with those of the 
conductivity tensor (related to the imaginary part of 
${\varepsilon}$) which is not generally valid as shown for instance for 
CdWO$_4$ \cite{Jellison2011} and Ga$_2$O$_3$ \cite{Sturm2015}. To overcome this 
problem, Höfer \textit{et al.} \cite{Hoefer2013,Hoefer2014} used for the 
infrared spectral range a model, developed earlier by Emslie \textit{et al.} 
\cite{Emslie1983}, which consists of a sum of damped Lorentz 
oscillators individually aligned to the axes of their respective dipole 
moments. For phonons, these axes are related to the atomic elongations and thus 
to some extent to the crystallographic axes. Further, their dissipative 
spectral range is usually narrow. Thus the question arises if such a model also can be applied to spectrally widespread excitations like electronic band-band 
transitions, which consist of numbers of individual 
dipoles whose axes are connected to overlapping atomic orbitals of 
various symmetry and therefore not necessarily coincide with crystallographic 
directions. Further the  density of states (DOS) of the electronic 
band structure is distributed within a wide energy range in a complex manner 
causing non-symmetric line-shapes of the imaginary part of the dielectric 
function which spectrally overlap for different contributions and directions.

Here we demonstrate that the sketched approach is generally valid for all kinds of 
excitations. We demonstrate this exemplarily for monoclinic Ga$_2$O$_3$ ($\beta$-phase) single crystals and thin films in the spectral range 
from infrared to vacuum ultraviolet. We show that this model provides 
a deep insight in electronic properties of the materials: Comparing the 
directions of the electronic polarizabilities obtained by modeling the experimental ellipsometry data using lineshape MDF to the real space atomic arrangement in the crystal and considering theoretical calculated electron density distribution as well as orbital-resolved DOS, allows us to assign the observed transitions to individual orbitals.

The paper is organized as follows: In Sec.~\ref{sec:df}, we discuss at first the dielectric tensor for all crystal symmetries and its composition. After that we demonstrate its applicability to the case of $\beta$-Ga$_2$O$_3$ single crystals in the infrared and ultraviolet spectral range. Finally, we show by means of a practically relevant $\beta$-Ga$_2$O$_3$ thin film which exhibit rotation domains that the approach of using directed transitions explains the effective uniaxial properties of the film and enhances the sensitivity to the out-of plane component of the dielectric tensor.

\section{Dielectric function}
\label{sec:df}

\label{sec:dielectric_function}
The optical response of a material is determined in first order by dipole 
excitations, e.g. 
optical phonons, electronic band-band transitions or excitons which in sum  
are represented by the dielectric function. For isotropic materials, the 
corresponding dipole moment or polarization direction of each excitation is 
macroscopically equally distributed in all spatial directions, resulting in an 
isotropic dielectric function, i.e. it is a scalar written as 
\begin{equation}
	{\varepsilon} = \mathds{1} + \sum_{i=1}^N {\varepsilon}_i\,,
	\label{eq:df_isotropic_sum}
\end{equation}
with $N$ being the number of excitations/oscillators. The situation 
changes for materials with crystal structure symmetry lower than the cubic one. In 
this case the excitations generally differ between the crystallographic 
directions in energy, amplitude, broadening, and even in the spatial direction 
of their dipole moment, and thus the DF is a tensor 
(Eq.~\eqref{eq:df_tensor_general}).

Let $\varepsilon_i'$ being the dielectric response of the $i^{th}$ excitation 
and the coordinate system is chosen (without loosing generality) in such way 
that the polarization direction is along the $x$-axis. The only non-zero component is then given by $\varepsilon_{xx}$, i.e. 
${\varepsilon}_{i,xx}'  \neq {\varepsilon}_{i,mn}' = 0$. 
However, the polarization direction of the excitation and the experimental 
coordinate system do not coincide with each other in general and a coordinate 
transformation has to be performed, independently for each transition. The 
entire dielectric tensor then can be expressed by   
\begin{equation}
{\varepsilon} = \mathds{1} + \sum_{i=1}^N R(\phi_i, \theta_i){\varepsilon}'_i R^{-1}(\phi_i, \theta_i)\,,
\label{eq:df_anisotropic_sum}
\end{equation}
with $\phi_i$ and $\theta_i$ being Euler angles, which are in general different 
for each excitation, and $R$ being the rotation matrix. The advantage of this expression 
is that the components of the resultant dielectric tensor in the Cartesian 
coordinate system are not independent of each other but rather composed of the 
respective projected part of the excitation's line-shape function according to 
the directions of their individual dipole moment. For the entire dielectric 
function it follows that, due to the finite broadening of each excitation and 
by considering Kramers-Kronig relation, the orientation of the 
principal tensor axes of the real and imaginary parts differ from each other 
as it is well known and observed in experiments e.g. for CdWO$_4$ 
\cite{Jellison2011} and Ga$_2$O$_3$ \cite{Sturm2015}.

Equation~\eqref{eq:df_anisotropic_sum} represents the general case which has to be used for triclinic crystals and can be 
simplified depending on the crystal symmetry. Crystals with monoclinic 
structure exhibit one symmetry axes, representing a $C_2$ rotation axis or the normal of a mirror plane (or both), which we identify in the following 
with the $y$-direction. The plane perpendicular to $y$, the $x$-$z$-plane, 
reveals no symmetry which defines a Cartesian coordinate system preferentially.
Therefore, from symmetry arguments,  considering dipoles 
polarized either along $y$ or in the $x$-$z$-plane, one can 
simplify Eq.~\eqref{eq:df_anisotropic_sum} to
\begin{equation}
 \varepsilon = \mathds{1}+ \sum_{i = 0}^{N_\text{y}} {\varepsilon}_{i,\text{y}}
	 + \sum_{j = 0}^{N_\text{xz}} R(\phi_j)\varepsilon'_{j,\text{xz}} R(\phi_j)^{-1}\,,
	 \label{eq:df_ga2o3_tensor}
\end{equation}
with $\varepsilon_{i,\text{y}}$ and $\varepsilon'_{j,\text{xz}}$ being the 
contribution of the respective directions. $N_\text{y}$ and $N_\text{xz}$ 
represent the number of excitations with the corresponding polarization 
directions and as $\phi$ we define the angle between the polarization 
direction and the $x$-axes within the $x$-$z$-plane. This leads to the well 
known form of the dielectric tensor given by 
\begin{equation}
	{\varepsilon} =
	\begin{pmatrix}
		{\varepsilon}_\text{xx} & 0 & {\varepsilon}_\text{xz}\\
		0 & {\varepsilon}_\text{yy} & 0 \\
		{\varepsilon}_\text{xz} & 0 & {\varepsilon}_\text{zz}
	 \end{pmatrix}\,.
	 \label{eq:tensor_monoclin}
\end{equation}

A further simplification can be made for orthorhombic materials containing 
three orthogonal twofold rotation symmetry axis, leading to
\begin{equation}
	 \varepsilon = \mathds{1}+ \sum_{i = 0}^{N_\text{x}} {\varepsilon}_{i,\text{x}}
	 +  \sum_{j = 0}^{N_\text{y}} {\varepsilon}_{j,\text{y}} +  
\sum_{k = 0}^{N_\text{z}} {\varepsilon}_{k,\text{z}}\,,
	\label{eq:df_orthrhombic}
\end{equation}
a dielectric function tensor which contains only diagonal elements. In the case 
of uniaxial materials, e.g. those with a hexagonal symmetry, 
${\varepsilon}_{i,\text{x}} = {\varepsilon}_{i,\text{y}}$ and $N_x = 
N_y$ holds. For isotropic material, the numbers of oscillators in all three 
directions is the same and therefore the dielectric tensor reduces to the scalar 
given by Eq.~\eqref{eq:df_isotropic_sum}.

For practical application Eq.~\eqref{eq:df_anisotropic_sum} has to be further 
modified. The real and imaginary parts of the dielectric function are connected 
with each other by the Kramers-Kronig relation. Contributions of 
excitations at energies higher than the investigated spectral range to the real 
part of the DF have to be considered. These contributions are usually described 
by a pole function. In the case presented here, this means that the identity in 
Eq.~\eqref{eq:df_anisotropic_sum} has to be replaced by a real valued tensor with 
the form given by the corresponding crystal structure where each component is represented by a pole function. 

\section{Experimental}
\label{sec:exp}
By using the approach presented in Sec.~\ref{sec:dielectric_function} and 
line-shape MDFs, the parametrised dielectric function of $\beta$-Ga$_2$O$_3$ 
bulk single crystals and thin films was determined in the mid-infrared up to the 
vacuum-ultraviolet spectral range by means of 
generalized spectroscopic ellipsometry.

Ga$_2$O$_3$ crystallizes at ambient conditions in monoclinic crystal 
structure, the so-called $\beta$-phase (Fig. \ref{fig_polarisation-crystal}). 
The  angle  
between the non-orthogonal $a$- and $c$-axis is $\beta=103.7^\circ$ 
\cite{Geller1960} resulting in a non-vanishing off-diagonal element of the 
dielectric tensor within the Cartesian coordinate system \cite{Sturm2015, 
Schubert2015}. We investigated two single side polished 
bulk single crystals from Tamura Corporation with $(010)$ and $(\bar{2}01)$ 
orientation, allowing access to all components of the dielectric tensor. 
X-ray diffraction (XRD) measurements does not reveal any hints for the presence 
of 
rotation domains, twins or in-plane domains. More details can 
be found in Ref. \onlinecite{Sturm2015}. The thin film was deposited on a 
$c$-plane 
oriented sapphire substrate by means of 
pulsed laser deposition (PLD) at $T \approx 730^\circ\text{C}$. After 
deposition, the sample was annealed for $5\,\mathrm{min}$ at $T \approx 730^\circ\text{C}$ and a oxygen partial pressure of $p_{\text{O}_2} = 800\,\mathrm{mbar}$. XRD measurements confirm the 
monoclinic crystal structure of the film and the surface orientation was 
determined to be  $(\bar{2}01)$. In contrast to the bulk single crystals, six 
rotation domains are observed which are rotated against each other by an 
angle of $60^\circ$. \cite{Splith2011} In contrast to bulk single crystals which reveal a smooth surface without atomic steps, the surface roughness of the thin film was determined to be $R_s \approx 5\,\mathrm{nm}$


In spectroscopic ellipsometry, the change of the polarization 
state of light after interaction with a sample is determined. In the general 
case, this is expressed by means of the $4 \times 4 $ Mueller matrix (MM, $\bf 
M$) which connects   the Stokes vectors of the incident (reflected) 
light $S_\text{in}$ ($S_\text{ref}$) by $S_\text{ref} = \bf{M}$\,$S_\text{in}$. 
In the special case where no energy transfer between orthogonal 
polarization eigenmodes of the probe light takes place, like for isotropic samples or optically uniaxial samples with the optical axis pointing along the surface normal (as the case for the 
thin film, cf. Sec. 
\ref{sec:thinfilm}), the change of the polarization state is expressed by the 
ratio of the complex reflection coefficients, i.e.  $\rho = \tilde{r}_\text{p} / \tilde{r}_\text{s}$. The index represents the polarization of the light polarized parallel ($p$) or perpendicular ($s$), respectively, to the plane of incidence which is spanned by the surface normal and the light beams propagation direction. 

For the determination of the DF, the experimental data are analyzed by 
transfer-matrix calculations considering a layer stack model. For the bulk 
single crystals, the model consists of a semi-infinite substrate (Ga$_2$O$_3$ 
itself) and a surface layer accounting for some roughness or 
contaminations. For the infrared spectral range the surface layer can be 
neglected. For the thin film the model consists of a $c$-oriented sapphire 
substrate, the Ga$_2$O$_3$ thin film layer and the surface layer. The dielectric 
function of sapphire was taken from the literature \cite{Yao1999}. The surface 
layer was modelled using an effective medium approximation (EMA) 
\cite{Bruggeman35} mixing the DF of Ga$_2$O$_3$ and void by $50\% : 50 \%$ for the bulk single crystals.\cite{Sturm2015} For the thin film this fraction was chosen as parameter and the best match between experiment and calculated spectra was obtained for $80\% : 20 \%$. In the following we choose our coordinate system in such way, that $\hat{e}_x \parallel a$-axis, $\hat{e}_y \parallel b$-axis and $\hat{e}_z = \hat{e}_x \times \hat{e}_y$.

\section{Bulk Single Crystals}
\label{sec:bulk_crystal}

\subsection{Infrared spectral range}

\begin{figure}
	\includegraphics[width = 0.95\columnwidth]{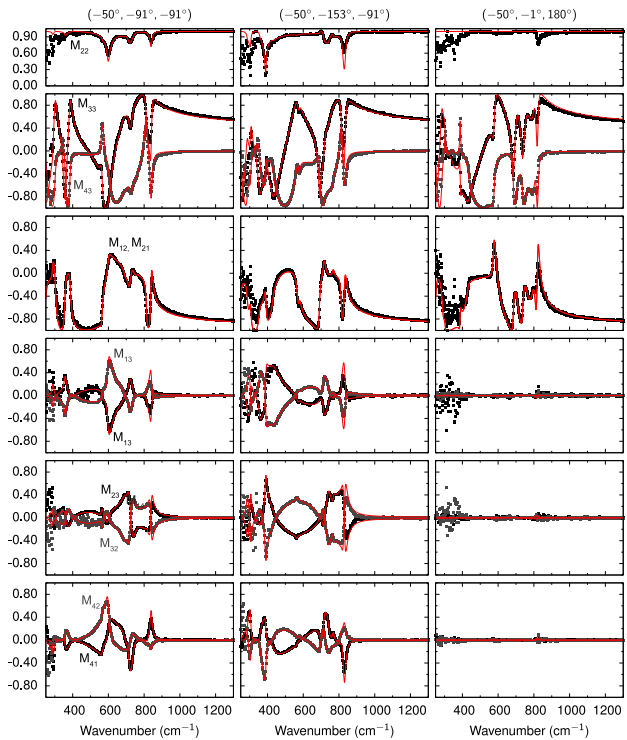}
	\caption{Experimental (symbols) and calculated (lines) spectra of the MM elements of a $\beta$-Ga$_2$O$_3$ bulk single crystal for an angle of incidence of $70^\circ$. The corresponding orientation of the crystal is given by the Euler angles on top of each column in the $yzx$ notation.}
	\label{fig:mm_ir}
\end{figure}

The MM in the infrared spectral ($250-1300\,\mathrm{cm}^{-1}$ ($31-161\,\mathrm{meV}$)) range was measured at angles 
of incidence of $30^\circ$, $50^\circ$ and $70^\circ$ for different in-plane 
rotations, i.e. rotating the crystal around its surface normal by  $30^\circ$, 
$60^\circ$ and $90^\circ$. For selected orientations the recorded spectra are 
shown in Fig.~\ref{fig:mm_ir}. The non-vanishing block-off-diagonal elements of 
the MM demonstrate the optically anisotropic character of the sample.

The dielectric function in the infrared spectral range is determined by phonon 
and free charge carrier oscillations. The bulk single crystals are not intentionally doped, the latter contribution can be neglected for the spectral 
range investigated here. 
Therefore, only phonons have to be considered and their contribution is described by Lorentzian oscillators: 
\cite{YUCardona2010}

\begin{equation}
  \varepsilon(E) = \frac{A \gamma E_0}{E_0^2 - E^2 - i\gamma E}\,,
\end{equation}

with $A$, $E_0$ and $\gamma$ being the amplitude, energy and broadening of the 
phonon mode, respectively. The calculated MM spectra are shown in 
Fig.~\ref{fig:mm_ir} as red solid lines yielding good agreement with the 
experimental ones. Note, that a similarly good match is obtained by using a 
Kramers-Kronig consistent numerical analysis and consider the four components of the DF (Eq.~\eqref{eq:tensor_monoclin}) to be independent from each other. 

In the investigated spectral range, 9 of the totally 12 optical infrared active phonon modes are observable. Their properties are summarized in Tab.~\ref{tab:phonon_frequency}. For the modes which have a dipole moment in the $a$-$c$ plane ($B_u$-symmetry) the 
polarization direction with respect to the $a$-axis is given by the angle 
$\phi$, which was found to differ for each phonon mode. This is also in agreement with the results recently reported by Schubert \textit{et al.}\cite{Schubert2015} The phonon mode $B_u^{(4)}$ was not observable in our experiment. This can be attributed to the 
weak sensitivity to this mode caused by its low amplitude which is predicted 
by ab-initio calculations (see below) and to the pronounced noise  caused by the 
low sensitivity of the detector of our setup in this 
spectral range. Further, for the mode $B_u^{(3)}$ 
only the frequency is given since also the large noise in this spectral range 
and the probable  spectral overlap with $B_u^{(4)}$ prohibit the determination 
of its dipole direction.

\begin{table*}
	\begin{tabular}{|ccccccccc||cc|ccc|}
	\hline \hline
 & $A$ & $f_\text{exp}/f_{0,e}$ & $f_\text{calc} / f_{0,c}$ & $\gamma$ & $\phi_\text{exp}$ & $\phi_\text{theo}$ & $E_{0,\text{exp}}$ & $E_{0,\text{theo}}$ & \multicolumn{2}{c|}{$E_{0,\text{exp}}$} & \multicolumn{3}{c|}{$E_{0,\text{theo}}$}  \\
 & \multicolumn{1}{p{1cm}}{} & & & $(\mathrm{cm}^{-1})$ & $(^\circ)$ & $(^\circ)$ & $(\mathrm{cm}^{-1})$ & $(\mathrm{cm}^{-1})$ & $(\mathrm{cm}^{-1})$ & $(\mathrm{cm}^{-1})$ & $(\mathrm{cm}^{-1})$ & $(\mathrm{cm}^{-1})$ & $(\mathrm{cm}^{-1})$ \\ \hline \hline 
$A_u^{(1)}$   &  -  & - & 0.01 &  -  &  -  & - &  -  &  160.7  & 155 &  154.8  & 155 &  155.7  & 141.6 \\
$B_u^{(1)}$   &  -  & - & 0.41 &  -  &  -  & 101 &  -  &  224.3  & 250 &  213.7  & 216 &  202.4  &  187.5 \\
$B_u^{(2)}$   & 92 & - & 0.33 & 8 &  -  & 176 & 253 &  267.3  & 290 &  262.3  & 300 &  260.4  &  251.6 \\
$B_u^{(3)}$   &  -  & - & 0.03 &  -  &  -  & 39 &  -  &  281.9  & 310 &  279.1  & 337 &  289.7  &  265.3 \\ 
$A_u^{(2)}$   & 51 &  - & 0.50 & 21 &  -  & - & 295 &  300.5  &  n.o.  &  296.6  & 352 &  327.5  &  296.2 \\
$B_u^{(4)}$   & 86 & 0.20 & 0.16 & 4 & 166 & 173 & 357 &  361.0  & 375 &  356.8  & 374 &  365.8  &  343.6 \\ 
$B_u^{(5)}$   & 82 & 0.96 & 0.96 & 17 & 46 & 47 & 430 &  434.2  & 455 &  432.5  & 500 &  446.8  &  410.5 \\
$A_u^{(3)}$   & 83 & 0.77 & 0.78 & 13 &  -  &  & 447 &  447.0  & 525 &  448.6  & 526 &  475.7  &  383.5 \\
$B_u^{(6)}$   & 73 & 1.00 & 1.00 & 15 & 128 & 130 & 572 &  560.8  & 640 &  572.5  & 626 &  589.9  &  574.3 \\
$A_u^{(4)}$   & 73 & 0.39 & 0.43 & 5 &  -  &  & 662 &  665.8  & 668 &  663.2  & 656 &  678.4  &  647.9 \\
$B_u^{(7)}$   & 32 & 0.25 & 0.23 & 7 & 28 & 0 & 691 &  692.5  & 692 &  692.4  & 720 &  705.8  &  672.6 \\
$B_u^{(8)}$   & 10 & 0.13 & 0.13 & 11 & 74 & 76 & 743 &  742.5  & 731 &  743.5  & 760 &  753.8  &  741.6 \\ \hline 

& \multicolumn{8}{c||}{this work} & Ref.~\onlinecite{Dohy1982} & Ref.~\onlinecite{Schubert2015} & Ref.~\onlinecite{Dohy1982} & Ref.~\onlinecite{Schubert2015} & Ref.~\onlinecite{Liu2007} \\ \hline \hline
	\end{tabular}
	\caption{Amplitudes ($A$), oscillator strength ($f_\text{exp}$),  damping parameters ($\gamma$) and energies ($E_{0,\text{exp}}$) of the phonon modes. The angle $\phi_\text{exp}$ represents the determined angle between the $a$-axes and the direction of the dipole moment in the $x$-$z$-plane. The phonon energy, oscillator strength and direction of the dipole from ab-initio calculations are given by $E_{0,\text{calc}}$ $f_\text{clc}$ and $\phi_\text{calc}$, respectively. For comparison the experimentally determined as well as calculated oscillator strength was normalized to those of the $B_u^{(6)}$ mode.}
	\label{tab:phonon_frequency}

\end{table*}

For comparison we calculated the phonon modes by ab-initio calculations based on the B3LYP hybrid functional approach implemented in the CRYSTAL14 code \cite{Kranert2015}. Thereby we used the basis set of Pandey \textit{et 
al.} \cite{Pandey1994} for gallium and of Valenzano \textit{et al.} 
\cite{Valenzano2006} for oxygen, which we slightly modified, and 150 $k$-points 
in the irreducible Brillouin zone. The truncation criteria defined by CRYSTAL14 
code given by five tolerance set to 8,8,8,8, and 16 for our calculations were 
used for the Coulomb and exchange infinite sums. Further we used a tolerance of 
the energy convergence of $10^{-11}$ Hartree. All input parameters and 
calculation
conditions can be found in Ref. \onlinecite{Kranert2015}. The calculated 
lattice parameters are $a=1.2336\,\mathrm{nm}$, 
$b=0.3078\,\mathrm{nm}$ and $c=0.5864\,\mathrm{nm}$, in reasonable 
agreement with those reported in the literature \cite{Geller1960}. The 
corresponding phonon mode energies, oscillator strength and the direction of the dipoles are also given in Tab.~\ref{tab:phonon_frequency} and are in excellent agreement with those determined by ellipsometry. The excellent agreement is not restricted to the infrared active phonon modes but is also obtained for the Raman active modes \cite{Kranert2015}.

\subsection{Ultraviolet spectral range}

The numeric DF in the UV spectral range was recently reported by us, obtained 
by using a Kramers-Kronig consistent numerical analysis \cite{Sturm2015}. 
In order to extract the properties of the contributing electronic transitions, 
e.g. energy and electronic orbitals involved, and to demonstrate the 
universal applicability of Eq.~\eqref{eq:df_anisotropic_sum} for electronic transitions we analysed the contribution of each transition to the entire DF by using line-shape model 
dielectric functions. Symmetry consideration and band structure 
properties\cite{Sturm2015} 
yield that the transitions are polarized either along the $y$-axes or within 
the $x$-$z$-plane. Thus the DF can be written as in 
Eq.~\eqref{eq:df_ga2o3_tensor} with a set of excitonic transitions and 
Gau{\ss}ian oscillators. We have been shown by density functional 
theory calculations combined with many-body perturbation theory including 
quasiparticle and excitonic effects \cite{Sturm2015}, that the DF in the 
spectral range from the fundamental absorption edge on up to some eV higher are 
dominated by excitonic correlation effects. Thus, several excitonic 
contributions have been considered in modeling and were 
described by a model dielectric function developed by C. Tanguy for Wannier 
excitons taking into account bound and unbound states. 
\cite{Tanguy1995,Tanguy1996,Tanguy1999}  The contribution of weakly 
pronounced band-band-transitions where summarized by using a Gaussian 
oscillator. A further Gaussian oscillator was included to consider 
contributions of transitions at energies higher than the investigated spectral 
range due to their spectral broadening. These contributions together with the pole function were considered for 
each dielectric tensor component independently because they may originate from 
different transitions.  

\begin{figure}
	\includegraphics[width = .99\columnwidth]{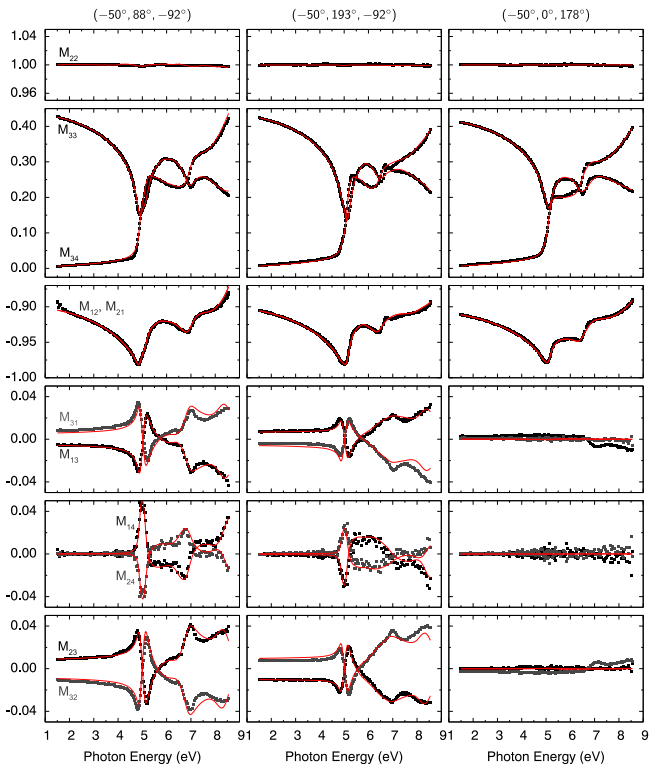}
	\caption{Experimental (symbols) and calculated (lines) spectra of the 
MM elements of a $\beta$-Ga$_2$O$_3$ bulk single crystal for an angle of incidence of $70^\circ$. 
The corresponding orientation of the crystal is given by the Euler angles on top of each column in the $yzx$ notation.} 
	\label{fig:mm_spectra_uv}	
\end{figure}

The experimentally recorded and the calculated spectra of the  MM elements are shown for selected 
orientations in Fig.~\ref{fig:mm_spectra_uv}, yielding good agreement. The 
difference between the experimental and the calculated spectra for energies $E > 7\,\mathrm{eV}$ was also observed by using the above mentioned numerical Kramers-Kronig consistent analysis and might be caused by the limitation of the used approach for the description of the surface layer.\cite{Sturm2015} This can be attributed to the fact that the sensitivity to this layer is strongly enhanced in this spectral range due to the enhanced absorption and therefore reduced penetration depth.

The parameters of the best-match MDF are summarized 
in Tab.~\ref{tab:transition_energies_part1} and \ref{tab:transition_energies_part2}. We extracted a exciton binding energy 
of about $E_\mathrm{X}^{b} = 270\,\mathrm{meV}$ for all contributions. Note 
that we considered the same exciton binding energy for all excitonic 
transitions because of the strong correlation between energy of 
the fundamental bound state and the corresponding binding energy.

\begin{table}
	\begin{tabular}{|ccccccc|}
	\hline \hline
	label & direction & Type  & $A$ & $E$ & $\gamma$ & $\phi$ \\ 
	& & & & (eV) & (meV) & ($^\circ$) \\ \hline \hline 
	$X_1$ & $a$-$c$ & Exciton & 15.0 & 4.88 & 70 & 110 \\
	$X_2$ & $a$-$c$ & Exciton & 18.0 & 5.10 & 800 & 17 \\
	$X_3$ & $a$-$c$ & Exciton & 14.9 & 6.41 & 210 & 41 \\
	$X_4$ & $a$-$c$ & Exciton & 28.0 & 6.89 & 190 & 121 \\
	$G_1$ & $a$-$c$ & Gauss & 0.27 & 6.14 & 1,343 & 124 \\ \hline 
	$X^b_1$  & $b$ & Exciton & 8.3 & 5.41 & 75 &  \\
	$X^b_2$ & $b$ & Exciton & 20.1 & 5.75 & 139 & \\
	$X^b_3$ & $b$ & Exciton & 7.0 & 6.93 & 253 &  \\ \hline \hline
	\end{tabular}
	\caption{Parameters of the UV model dielectric functions for the observed transitions within the investigated spectral range. The angle $\phi$ represents the orientation of the dipole moment in the $a$-$c$-plane with respect to the $x$-axis. }
	\label{tab:transition_energies_part1}
\end{table}

\begin{table}
	\begin{tabular}{|c|ccc|cc|c|}
	\hline \hline
	& \multicolumn{3}{c|}{Gauss} & \multicolumn{2}{c|}{Pole} &  $\varepsilon_\infty$\\
	& $A$ & $E$ & $\gamma$ & $E$ & A & \\
	&  & (eV) & (eV) & (eV) & A &  \\ \hline \hline
	$\varepsilon_{xx}$ & 2.64 & 9.69 & 2.7 & 200.8 & 15.5 & 0.907 \\
	$\varepsilon_{yy}$ & 1.81 & 9.78 & 3.8 & 52.7 & 10.5 & 1.392 \\
	$\varepsilon_{zz}$ & 1.84 & 8.99 & 1.7 & 91.5 & 11.9 & 1.126 \\
	$\varepsilon_{xz}$ & 0.26 & 8.49 & 0.5 &  &  & -0.086 \\ \hline \hline 
	\end{tabular}
	\caption{Parameters of the UV model dielectric function describing the contributions of the high energy transition to the dielectric function in the investigated spectral range.}
	\label{tab:transition_energies_part2}
\end{table}

\begin{figure*}
	\includegraphics[width = .95\textwidth]{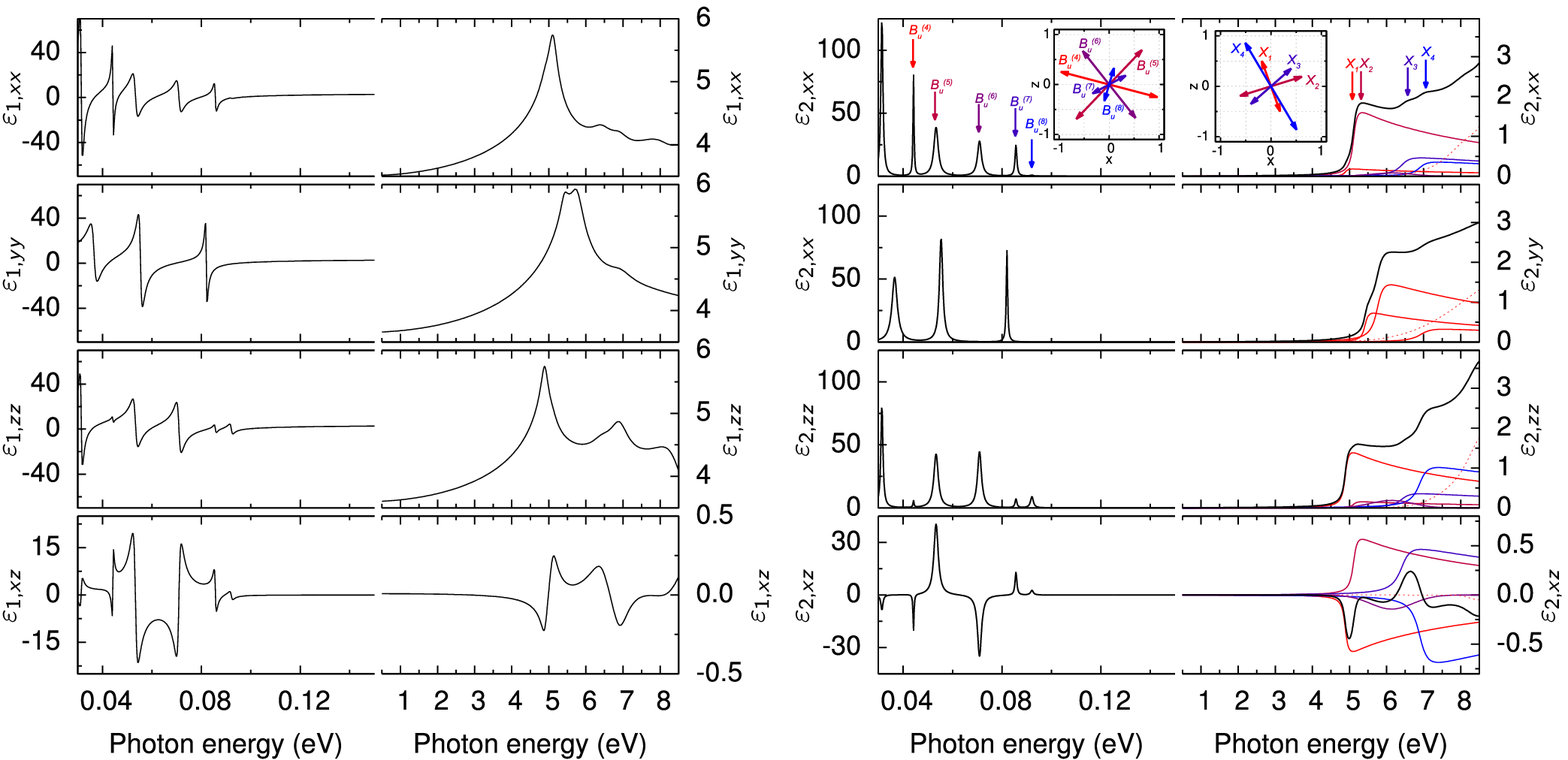}
	\caption{Dielectric function (black solid line) of a $\beta$-Ga$_2$O$_3$ bulk single crystal in the infrared and UV spectral range. The red solid lines represent the excitonic contribution in the investigated UV spectral range whereas the red dashed lines represent the contribution of the high-energy contributions. The arrows in the insets depict the orientation of the corresponding dipole moment and their relative amplitude ratio.} 
	\label{fig:df_complete}	
\end{figure*}

The dispersion of the tensor elements for the entire investigated spectral range 
is shown in Fig.~\ref{fig:df_complete}. The contributions of excitonic 
transitions to $\varepsilon_2$ are shown as 
red solid lines. The orientation of the corresponding dipole moments in the 
$x$-$z$-plane is indicated by the arrows in the inset. In agreement with our 
theoretical calculations and the numeric MDF,\cite{Sturm2015} 
the two energetically lowest transitions (labeled as $X_1$ and $X_2$) are strongly 
polarized along the $x$- and $z$-direction, respectively. At higher energies, there are transitions along $y$-axis ($b$-axis) and within the $x$-$z$-plane ($a$-$c$-plane).

\begin{figure*}
  \includegraphics[width = .7\textwidth]{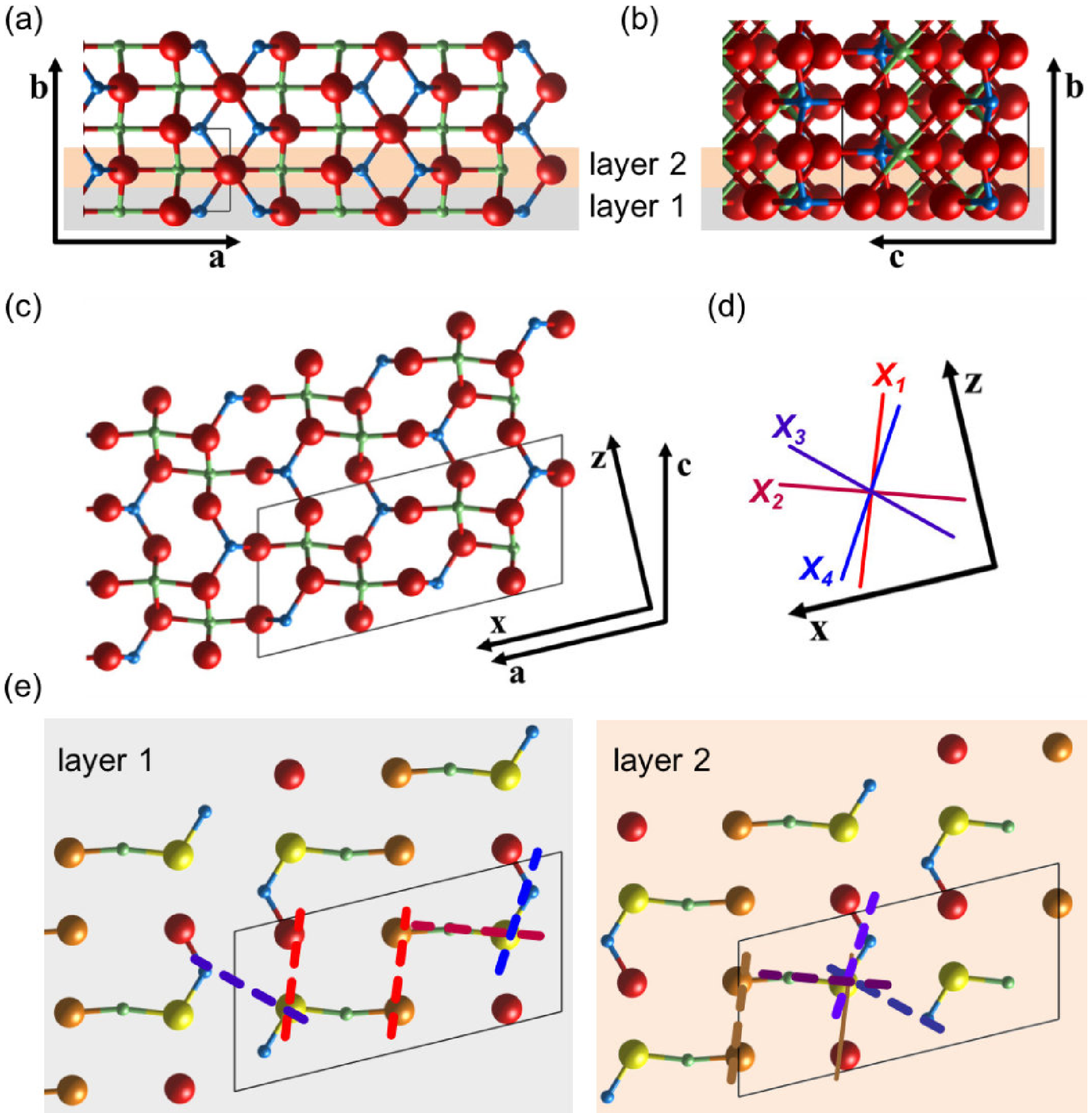}
  \caption{(a-c): Schematic representation of projections of the 
crystal structure of $\beta$-Ga$_2$O$_3$ into the $b$-$a$-plane (a), 
$b$-$c$-plane (b) and $c$-$a$-plane (c). The unit cell is 
indicated by the black framed boxes. Bonds are indicated by lines between 
the atoms. The respective Cartesian coordinates $x$ and $z$ are indicated, $y$ 
points along $b$. The tetrahedrally coordinated Ga(I) atoms are shown in blue 
and the octahedrally coordinated Ga(II) are shown in green. The oxygen atoms 
are marked in red. (d) The directions of the dipole moments within the 
$c$-$a$-/$x$-$z$-plane are indicated for the transitions $X_1$ \dots $X_4$ at 
the left side of the middle row. (e) Sub-layers of the $c$-$a$-plane (left: layer 1, right: layer 2) as 
indicated also in the upper row. Here, the oxygen atoms at different lattice 
sites are highlighted by colours as O(I) red, O(II) orange and O(III) yellow 
(see also text). The dashed coloured lines relate the dipole directions of 
the transitions $X_1$ \dots $X_4$ to atomic bonds within the crystal structure. 
Please note that only one example is shown for each different transition. (Images created by VESTA (Ref. \onlinecite{Vesta}).)}
  \label{fig_polarisation-crystal}
\end{figure*}

Based on calculated charge distribution \cite{Litimein2009} and atomic arrangement within the $x$-$z$-plane ($a$-$c$-plane), we relate the directions of the dipole moments of all four pronounced excitonic excitations within this plane ($X_1$ \dots $X_4$), as obtained from the ellipsometry model, to atomic bonds in the crystal structure as shown in Fig.~\ref{fig_polarisation-crystal}. %
For transitions along $y$, no direct assignment 
to individual orbitals was possible because of the complex distribution of 
atomic bonds. Please note that the uncertainty in the experimentally determined 
dipole moment directions amounts 
to up to 10$^{\circ}$, caused by the simplification due to the used model 
functions, which summarize spectrally over different individual transitions. As 
all these transitions reveal no contribution to the dielectric tensor component 
$\varepsilon_{\text{yy}}$, only bonds located solely within the sub-planes of 
the $x$-$z$-plane ($a$-$c$-plane) are considered (cf. Fig~\ref{fig_polarisation-crystal}).  It is found that all excitonic transitions but the first one, which appears to take place between oxygen atoms, 
are between differently coordinated gallium and oxygen.  In the following discussion we will use the nomenclature given by Geller \cite{Geller1960} and label the tetrahedrally and octahedrally coordinated Ga atoms as Ga(I) and Ga(II), respectively, while the three different sites of the oxygen atoms are labeled as O(I), O(II) and (OIII) (cf. Fig.~\ref{fig_polarisation-crystal}).
 
Band structure calculations reveal that the uppermost valence bands are dominated by oxygen 
$p$-orbitals, while the DOS of the lowest conduction bands is composed of 
almost equal contributions from Ga-$s$, O-$s$, and O-$p$ 
orbitals.\cite{Yamaguchi2004,Litimein2009,He2006} Thus, dipole allowed transitions can take place from O-$p$ orbitals to Ga-$s$ and O-$s$ orbitals. It turns out that the states near the conduction band minimum are preferentially determined by octahedrally coordinated Ga(II). \cite{Litimein2009} 
This is reflected by the assignment of the 
dipole directions to the atomic bonds in Fig. \ref{fig_polarisation-crystal}. It 
turns out that the transition $X_2$, almost directed along $x$ ($a$) involves O 
and Ga(II) and also reveals a high amplitude in the DF. Ga(II) is located 
between O(II) and O(III). But the dipole direction only fits to the bond 
Ga(II)-O(III), so it seems that transitions to Ga(II) states in the conduction 
band only appear when O(III) is involved and are not possible involving O(II). 
This can be understood considering the coordination of the O-atoms, which is 
higher (6 bonds) for O(II), suggesting the orbitals to be more $s$-like compared 
to O(III) (4 bonds) 
which dominate the DOS near the valence band maximum. %
The transitions $X_3$ and $X_4$ are assigned to take place between Ga(I) and O(III). The directions obtained from model analysis of the DF does not fit as good as for transition $X_2$, maybe caused in correlation effects due to spectral overlap of different contributions 
to the DF. 
Finally, transition $X_1$, directed almost 
along $c$, was assigned to take place either between O(I) and O(III) or between 
two O(II) atoms, or both. While the first possibility involves differently 
coordinated atoms suggesting dipole allowed transitions between $p$- and 
$s$-like orbitals, the second possibility involves only highly coordinated atoms 
($s$-like character) and thus should be dipole forbidden. The relatively high 
amplitude of this transition is not clear at first place, because following Ref. 
\onlinecite{Litimein2009}, the charge density between the involved atoms and 
also the DOS of the 
oxygen orbitals in the conduction band is predicted to be relatively weak. 

These results nicely demonstrate the potential of the used model approach for 
the dielectric tensor to gain deep insight into electronic properties of highly 
anisotropic materials.

\section{thin film}
\label{sec:thinfilm}

As mentioned above, the PLD grown $\beta$-Ga$_2$O$_3$ thin film exhibit  
$(\bar{2}01)$ surface orientation with 6 in-plane rotation domains, 
rotated by multiples of $60^\circ$. As their size is much smaller than the 
optically probed sample area of about $5 \times 8$\,mm$^2$, the measured 
optical response is determined by an average over these domains. For uniform 
distribution of these rotation domains, the effective dielectric function is 
given by 
\begin{align}\nonumber
	\varepsilon & = \sum_{i=1}^{6} R(\phi_i) 
		\varepsilon^\text{mono}
		R^{-1}(\phi_i) \\
		& =
	\begin{pmatrix}
		0.5(\varepsilon'_{xx} + \varepsilon_{yy}) & 0 & 0 \\
		0 & 0.5(\varepsilon_{xx} + \varepsilon_{yy}) & 0 \\
		0 & 0 & \varepsilon'_{zz} 
	\end{pmatrix}\,,
	\label{eq:df_film}
\end{align}

with $\phi = (i-1)\pi / 3 $ the rotation angle of the $i^\text{th}$ rotation domain ($i=1 \dots 6)$ and $R(\phi)$ being the rotation matrix around the surface normal. Equation~\eqref{eq:df_film} is similar to those of a uniaxial material with $\varepsilon_\perp = 0.5(\varepsilon'_{xx} + \varepsilon_{yy}) $ and $\varepsilon_\parallel = \varepsilon'_{zz}$ ($\perp$ and $\parallel$: perpendicular and parallel to the optical axis) with orientation of the effective optical axis along the surface normal. Note that $\varepsilon'_{xx}$ and $\varepsilon'_{zz}$ are the tensor components for the coordinate system with the $x$- and $z$ axis parallel and perpendicular to the sample surface, respectively.

For such samples with the sensitivity to $\varepsilon_\parallel$ is usually limited due to the high index of 
refraction of the investigated material resulting in a propagation direction of the wave within the sample with only very small angles off the optical axis. But there is a finite projection of the electro-magnetic field strength onto the optical axis and thus the optical response is determined by $\varepsilon_\perp$ and $\varepsilon_\parallel$ in any case, which have to be considered in order to obtain a physical meaningful dielectric function \cite{Shokhovets2010}. However, in contrast to a homogeneous uniaxial material, those effective $\varepsilon_\perp$ and $\varepsilon_\parallel$ are not independent from each other. As shown in Sec.~\ref{sec:dielectric_function} and demonstrated in Sec.~\ref{sec:bulk_crystal} the components $\varepsilon'_{xx}$ and $\varepsilon'_{zz}$ reflect the same transitions and are determined by the projection  $A'_{zz}/A'_{xx} = \sin^2\phi' / \cos^2\phi'$ of their amplitudes $A$ ($\phi'$ is the angle of the oscillation direction of the individual dipoles with respect to the sample surface). This offers in the present case more sensitivity for determination of the tensor component  $\varepsilon_\parallel$ as compared to homogeneous uniaxial materials.

\begin{figure}
	\centering
	\includegraphics[width = 0.99\columnwidth]{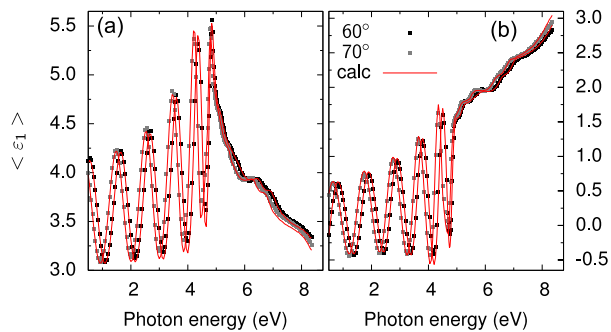}
	\caption{Real (a) and imaginary (b)  part of the thin film's pseudo 
dielectric function for angle of incidence 60$^{\circ}$ and 70$^{\circ}$. The 
experimental and calculated data are shown as symbols and red solid lines, 
respectively.}
	\label{fig:spectra_film}
\end{figure}

The uniaxial behaviour of the film with the optical axis parallel to the surface 
normal is reflected by vanishing off-diagonal elements of the MM. Therefore, 
standard ellipsometry is sufficient for measuring the full optical response 
(cf. Sec.~\ref{sec:exp}). The experimental data are shown in 
Fig.~\ref{fig:spectra_film} in terms of the pseudo dielectric 
function\cite{Fujiwara}  
\begin{align}\nonumber
<\varepsilon> &= <\varepsilon_1>+i<\varepsilon_2>  \\
&= \sin^2 \Phi \left[1+\tan^2 \Phi \left(\frac{1-\rho}{1+\rho}\right)^2\right]
\end{align} 
with angle of incidence $\Phi$. Below $E \approx 4.8\,\mathrm{eV}$ oscillations due to multiple reflection interferences caused by the interfaces within the sample are observed which vanish  with the onset of the absorption at higher energies.

\begin{figure}

	\centering
	\includegraphics[width = 0.99\columnwidth]{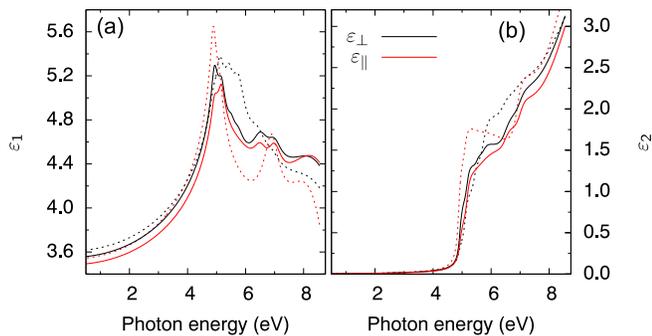}
	\caption{Real (a) and imaginary (b) part of the tensor components of the 
Ga$_2$O$_3$ thin film (solid lines). For comparison, the components calculated 
by Eq.~\eqref{eq:df_film} using the single crystal values are shown as 
dashed lines.}
	\label{fig:df_film}
\end{figure}

For the parametric model of the dielectric function of the thin film we used 
the same set of model dielectric functions as for the bulk single crystal. The 
calculated spectra are shown as red solid lines in Fig.~\ref{fig:spectra_film} 
and a good agreement between the experimental and calculated data is 
apparent. The tensor components of the dielectric function of the thin film are 
shown in Fig.~\ref{fig:df_film}. For comparison, the components calculated from 
DF of the bulk single crystal by using Eq.~\eqref{eq:df_film} are shown as 
dashed lines. For the thin film, we needed to adjust 
energies and amplitudes of the transitions and even the dipoles' orientation 
angles $\phi$ within the $x$-$z$-plane ($a$-$c$-plane). Compared to the DF 
of the single crystal a blue-shift of the transition energies up to 
$100\,\mathrm{meV}$ and a lowering of the oscillator strengths is observed for 
the thin film. The reduced oscillator strength in the investigated spectral 
range cannot explain the lowering of the real part of the dielectric constant 
and therewith of the index of refraction in the visible spectral range alone. 
Therefore, the reduced refractive index indicates also a reduced oscillator 
strength of the high energy transitions compared to the bulk single crystal.
We relate these changes of the DF properties compared to the bulk single 
crystal on the one hand to crystal imperfections typically lowering the 
oscillator strength of electronic transitions by dissipative processes. On the other hand 
also strain will be possibly present in the thin film, causing changes in the 
bond length and maybe also torsion of the unit cell causing different dipole 
moment orientations.

\section{Summary}
We have determined the dielectric function of $\beta$-Ga$_2$O$_3$ by using a 
generalized oscillator model taking into account the direction of the dipole 
moments for each transition. Within this model, the components of the 
dielectric tensor within the cartesian coordinate system are not independent 
from each other but are determined by the projection of the corresponding 
dipole direction. In doing so, we could determine the tensor components of the DF of
$\beta$-Ga$_2$O$_3$ bulk single crystals and thin films. By means of the determined direction of the dipoles we assign the involved orbitals for the observed transitions. For the thin film we showed that the presence of rotation domains leads to the formation of an effective uniaxial material. The 
sensitivity to the out-of-plane component of the dielectric tensor is enhanced 
compared to pure uniaxial materials since it is connected to the in-plane 
component. This allows a precise determination of this component even if the 
optical axis is perpendicular to the surface, which is relevant for applications in optoelectronics.

\acknowledgments
We thank Hannes Krauß, Vitaly Zviagin and Steffen Richter for the support of the ellipsometry measurements. This work was supported by the Deutsche Forschungsgemeinschaft within Sonderforschungsbereich 762 -  "Functionality of Oxide Interfaces". We also acknowledge financial support of the Austrian Fond zur F\"orderung der Wissenschaftlichen Forschung in the framework of SFB25 - "Infrared Optical Nanostructures".

\end{document}